\documentclass[letterpaper,10pt,twocolumn,final,conference,oneside]{IEEEtran}
\usepackage{array}
\usepackage{epsfig}
\usepackage{graphics}
\usepackage{graphicx}
\usepackage[english]{babel}
\usepackage[latin1]{inputenc}
\usepackage{amsmath}
\usepackage{amssymb}
\usepackage{booktabs}
\usepackage{floatrow}
\usepackage{subfigure}
\usepackage{float}
\usepackage{bm}
\usepackage{cite}
\usepackage{cases}
\usepackage{color,soul}
\usepackage{comment}
\usepackage{multicol}
\usepackage{multirow}
\usepackage{balance}
\usepackage{acronym}
\usepackage{hyperref}

\usepackage{tikz}

\usepackage{siunitx}
\DeclareSIUnit{\atm}{atm}
\DeclareSIUnit{\kWh}{kWh}
\DeclareSIUnit{\Ah}{Ah}

\restylefloat{table}
\acrodef{wgsr}[WGSR]{water gas shift reactor}
\acrodef{bess}[BESS]{battery energy storage system}
\acrodef{ess}[ESS]{energy storage system}
\acrodef{sps}[SPS]{shipboard power system}
\acrodef{ips}[microgrid]{microgrid}
\acrodef{pms}[PMS]{power management system}
\acrodef{pemfc}[PEMFC]{proton exchange membrane fuel cell}
\acrodef{fc}[FC]{fuel cell}
\acrodef{soc}[SOC]{state of charge}
\acrodef{kvl}[KVL]{Kirchhoff's voltage law}
\acrodef{kcl}[KCL]{Kirchhoff's current law}
\acrodef{cpl}[CPL]{constant power load}
\acrodef{cps}[CPS]{constant power source}
\acrodef{res}[RES]{renewable energy source}
\acrodef{aes}[AES]{all-electric ship}
\acrodef{s-sasc}[S-SASC]{small-signal asymptotic stability condition}

\IEEEoverridecommandlockouts

\newcommand\copyrighttext{%
  \footnotesize
  \centering\copyright~2021 IEEE. Personal use of this material is permitted. Permission from IEEE must be obtained for all other uses, in any current or future media, including reprinting/republishing this material for advertising or promotional purposes, creating new collective works, for resale or redistribution to servers or lists, or reuse of any copyrighted component of this work in other works.\\
  IEEE PES General Meeting 2021. \href{https://doi.org/10.1109/PESGM46819.2021.9638117}{10.1109/PESGM46819.2021.9638117}}
\newcommand\copyrightnotice{%
\begin{tikzpicture}[remember picture,overlay]
\node[anchor=south,yshift=0pt] at (current page.south) {\setlength{\fboxrule}{0pt}\fbox{\parbox{\dimexpr\textwidth-\fboxsep-\fboxrule\relax}{\copyrighttext}}};
\end{tikzpicture}%
}

\begin{document}

\title{Small-Signal Stability Analysis of a DC Shipboard Microgrid With Droop-Controlled Batteries and Constant Power Resources}

\author{%
 \IEEEauthorblockN{F. Conte, F. D'Agostino, S. Massucco, F. Silvestro}
 \IEEEauthorblockA{University of Genoa\\
    DITEN\\
    Via all'Opera Pia 11 A\\
    I-16145, Genova, Italy\\
    fabio.dagostino@unige.it}
 \and
  \IEEEauthorblockN{S. Grillo}
  \IEEEauthorblockA{Politecnico di Milano\\
    Dipartimento di Elettronica, Informazione e Bioingegneria\\
    p.zza Leonardo da Vinci, 32\\
    I-20133, Milano, Italy\\
    samuele.grillo@polimi.it}}

\IEEEaftertitletext{\copyrightnotice\vspace{0.2\baselineskip}}
\maketitle
\acused{ips}
\begin{abstract}
The presence of constant power loads (CPLs) in dc shipboard microgrids may lead to unstable conditions. The present work investigates the stability properties of dc microgrids where CPLs are fed by fuel cells (FCs), and energy storage systems (ESSs) equipped with voltage droop control. With respect to the previous literature, the dynamics of the duty cycles of the dc-dc converters implementing the droop regulation are considered. A mathematical model has been derived, and tuned to best mimic the behavior of the electrical representation implemented in DIgSILENT. Then the model is used to find the sufficient conditions for stability with respect to the droop coefficient, the dc-bus capacitor, and the inductances of the dc-dc converters.
\end{abstract}



\acresetall\acused{ips}
\section{Introduction}\label{sec:Introduction}

The growing concern on clean energy has widely interested transportation systems. This is impacting not only ground transportation means, but also marine systems. In the decades, the design criteria of vessels have been modified to give a more significant role to both the generation and the usage of electrical energy. As a matter of fact, not only has electric energy been used for feeding final users, but it has been exploited as a propulsion means too. This concept is widely known as \ac{aes}.

This has led to the installation on board of diverse technological solutions which had been left ashore in the past years. Given the fact that the most common configuration of \acp{aes} has traditional Diesel gensets, \acp{ess} have been recently introduced to increase both efficiency and performances of the shipboard power system. In addition to this, \acp{res} have been installed to mitigate polluting emissions. In such systems, the presence of \acp{ess} is also instrumental to reduce the randomness, and improve efficiency and reliability of \acp{res}.
A further step towards more environmentally-sustainable solutions is to increase the share of green power produced on-board. This goal could not be attained using only \acp{res}. Hence more stable and reliable sources of power have been considered. Among the alternatives, \acp{pemfc} have been regarded as an adequate means, as they guarantee many advantages, and zero local pollutant emissions, when fed by green hydrogen~\cite{gualeni}.

The introduction of such equipment brought on-board electrical power system to the limelight, and the idea of a dc \ac{ips} is emerging as an efficient and effective solution. In this configuration there are few dc busbars used to collect generated power and to distribute it to loads. A dc distribution system has many advantages as: i) both \acp{pemfc} and \acp{ess} have a native dc output, and ii) loads (both propulsion and hotel services) increase both their efficiency and controllability when interfaced by means of power converters---thus suggesting solutions with simpler dc-ac converters rather than ac-ac ones---. However, these converter-connected loads often behave as \acp{cpl}.

In a system with \acp{fc} and \acp{ess}, since the former sources do not display high dynamic performances, voltage regulation is guaranteed by \acp{ess} mainly through a droop control. The simultaneous presence of \acp{cpl} may impair the overall stability of the system, which has been studied in literature \cite{Anand2013,Herrera2017,Tahim2015,Su2018,Radwan2012,Liu2018}. However, one of the main assumptions in these works is that the duty cycles of the dc-dc converters of the \acp{ess} are supposed to be faster enough to neglect their influence in the dynamics.

Moving from this assumption, we started analyzing the influence of the duty cycles, thus including them in the states of system, and also adding a tunable delay (modeled as a filter, so as to minimize the complexity of the resulting equations) in order to take into account the natural delay that would be present in a real system. This value of the delay has been chosen as the minimum one which guarantee the matching between the model implemented in DIgSILENT and the mathematical model. A further analysis has been carried out, to assess the influence of the droop on stability and to find a design criterion for the sizing of the converters inductances, and of the dc-bus capacitor. In particular, given a certain value for the inductances, the study presented in this paper found the minimum value of the capacitor which guarantees stability.

The paper is structured as follows. In Section~II the components of the dc shipboard \ac{ips} are described. Section~III reports the model equations. By considering the duty cycles of the dc-dc converters as state variables, the system becomes nonlinear. The general layout of the linearized representation of the system is provided at the end of Section~III. In Section~IV the results of the stability analysis are reported. Finally, conclusions are drawn in Section~V.

\section{Shipboard Power System Architecture}\label{sec:SystemModel}

\begin{figure}[ht]
	\centering
    \includegraphics[scale=.25]{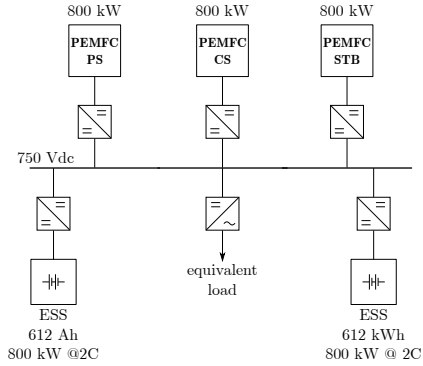}
    	\caption{One line diagram of the dc shipboard \ac{ips} under study.}
	\label{fig:Rete}
\end{figure}

The dc shipboard \ac{ips} under study comprises two \acp{ess} and three \acp{pemfc} connected at the main dc bus, as shown in Figure~\ref{fig:Rete}.

The converter-interfaced equivalent load represents the total ship power demand, including propulsion services, auxiliaries, heat ventilation and cooling, and hotel load.
\acp{fc} power conditioning system consists of a dc-dc boost converter, controlled to obtain a constant power source, to meet the poor dynamic behavior of \ac{pemfc}, as in~\cite{DAgostino:2019}. This control strategy allows the \ac{pemfc} generator to avoid power injection variations as a consequence of dc bus voltage fluctuations. On the other hand, each \ac{ess} is connected trough a bi-directional dc-dc converter designed to regulate the dc-bus voltage.

Each \ac{pemfc} source is a \SI{800}{\kilo\watt}-\SI{700}{\volt} generator, obtained with 4 parallel stacks composed by a series of 2 modules each. Each \ac{ess} considered for this study is a \SI{612}{\Ah}-\SI{690}{\volt} Lithium Nickel Manganese Cobalt Oxide (NCM) battery, composed by 9 parallel racks of 12 series trays. The main parameters of sources are reported in Table~\ref{tab:parms}.

\begin{table}[ht]
		\centering
		\caption{Power sources main parameters.}
		\label{tab:parms}
		{\small
		\begin{tabular}{|c|l|l|}
		\toprule
	\multicolumn{2}{|c|}{Parameter}            & \multicolumn{1}{c|}{Value} \\
	\midrule
\multirow{2}{*}{FC}  & nominal voltage     & \SI{700}{\volt}          \\
                     & nominal power       & \SI{800}{\kilo\watt}     \\ \midrule
\multirow{5}{*}{ESS} & nominal voltage     & \SI{690}{\volt}          \\
                     & nominal capacity    & \SI{612}{\Ah}            \\
                     & nominal power @2C   & \SI{800}{\kilo\watt}     \\
                     & max. power @4C      & \SI{1.71}{\mega\watt}    \\
                     & internal resistance & \SI{10}{\milli\ohm}      \\
		\bottomrule
		\end{tabular}
        }
\end{table}

A battery can be modeled as a voltage source in series with a resistance~\cite{Tremblay2007}. The battery voltage can be written applying the \ac{kvl}, as follows:
\begin{IEEEeqnarray}{rCl}
    v_{\rm B}& =& e_{\rm{B}}-R_{\rm B}^{\rm{int}}i_{\rm{B}}\label{eq:Vbat}
\end{IEEEeqnarray}
where $e_{\rm B}$ is the battery open-circuit voltage, which is a function of the battery state of charge \cite{DAgostino:2019}.

\subsection{Regulation and Control Strategy}
The power control strategy is based on the primary frequency regulation concepts, migrated to the voltage regulation of a single dc-bus power system. With the hypothesis that each power source is connected to the same electrical node, dc-bus voltage deviations from the reference value can be treated as the power unbalance. Thus, dc-bus voltage plays a role similar to that played by frequency in classical ac power systems.

The voltage regulation at the dc bus is provided by dc-dc converters of \acp{ess}, equipped with droop controllers. When the dc-bus voltage changes as a consequence of a load variation, the current of each regulating unit changes accordingly, following its droop characteristic.

The control action of the $j$-th converter is realized through the variation of the modulation index $\alpha_{{\rm B}_j}$, defined  as  the  ratio between  the  voltage  on  the  dc-bus  side and the input voltage at the battery side. PI regulators realize the primary dc-bus voltage control, as follows:
\begin{equation}\label{eq:delta_alpha}
    \Delta\alpha_{{\rm B}_j} =\left(k_{\rm P}+\frac{k_{\rm I}}{s}\right) \left(v^{\rm ref }_j - v\right),
\end{equation}
where $k_{\rm P}$ and $k_{\rm I}$, are the proportional and the integral gains respectively, $v$ is the actual dc-bus voltage (in p.u.), and $v^{\rm ref}_j$ is the reference voltage (in p.u.) calculated by the droop equation:
\begin{equation}\label{eq:droop}
v^{\rm ref}_j = v_{\rm 0}-D\left(i_{{\rm B}_j} - i_{0}\right)
\end{equation}
where $D$ is the droop coefficient (in p.u.), $v_{\rm 0}$ and $i_{\rm 0}$ are the p.u. idle values of voltage and current respectively, and $i_{\rm B}$ is the battery current (in p.u.). For the sake of simplicity, $k_{\rm P}$, $k_{\rm I}$, $i_{\rm 0}$, and $D$ are supposed to be the same for all \acp{ess}. Secondary voltage regulation is not considered here.

\section{Small-Signal System Modeling}\label{sec:SmallSignalModeling}
The Lyapunov method represents a powerful tool widely used for power system small-signal stability analysis. When a nonlinear autonomous system
\begin{IEEEeqnarray}{rCl}\label{eq:nlinsys}
    \dot{\boldsymbol{x}} = \boldsymbol{f}\left(\boldsymbol{x}\right)
\end{IEEEeqnarray}
where $\boldsymbol{x}$ is the system state, is linearized at the equilibrium point $\bar{\boldsymbol{x}}$ as
\begin{IEEEeqnarray}{rCl}\label{eq:linsys}
    \dot{\hat{\boldsymbol{x}}} = {\rm{\mathbf{A}}}\hat{\boldsymbol{x}},
\end{IEEEeqnarray}
the analysis of the eigenvalues of $\mathbf{A} = \left.J\boldsymbol{f}\right|_{\bar{\boldsymbol{x}}}$ (where $J$ is the Jacobian operator) allows to evaluate the local stability of the original system. In particular, if and only if all the eigenvalues of $\mathbf{A}$  have negative real parts, then it is possible to conclude that the actual system is locally asymptotically stable at the given equilibrium point.

\begin{figure}[ht]
	\centering
    \includegraphics[scale=.4]{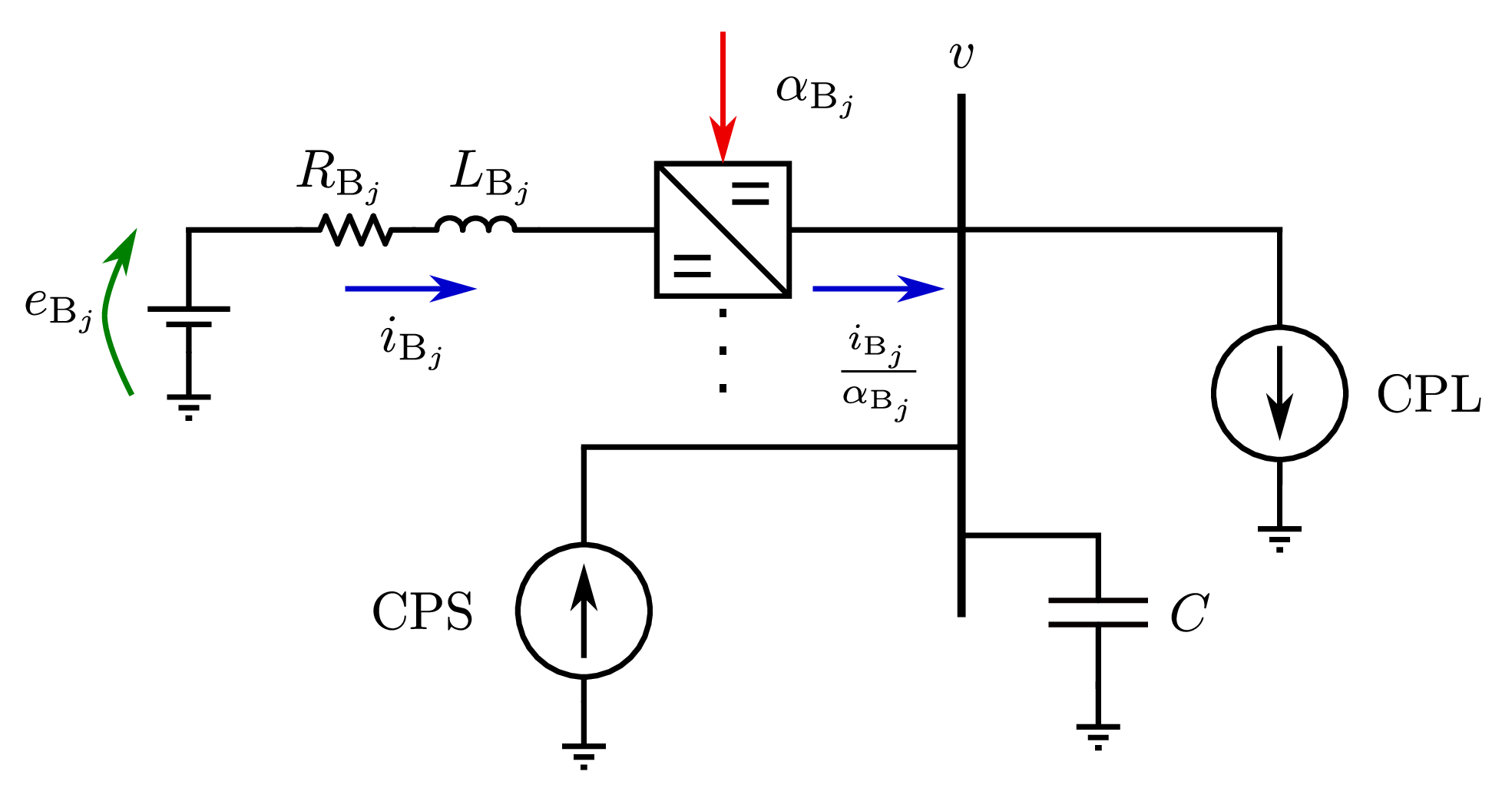}
    	\caption{Microgrid equivalent circuit.}
	\label{fig:dcgrid}
\end{figure}

The microgrid system is described by the voltage and current Kirchhoff equations of the circuit depicted in Fig.~\ref{fig:dcgrid}, where $e_{\rm {Bj}}$, $i_{\rm {Bj}}$, $R_{\rm {Bj}}$, $L_{\rm {Bj}}$ represent respectively the open circuit voltage, the output current, the sum of internal and filter resistances, and the filter inductance of the $j$-th battery source. The modulation index $\alpha_{\rm {Bj}}$ of dc-dc $j$-th converter allows the regulation of the main control variable, which is the voltage magnitude $v$ at the dc bus. The capacitance filter is indicated as $C$, while the equivalent generator representing the fuel cells, controlled as \acp{cps}, and the total \ac{cpl} are indicated as \ac{cps} and \ac{cpl}.

When we started with the system modeling, we initially considered modulation indices of batteries converters as constant quantities. This first setup allowed us to catch the inherent stability characteristics of the circuit, in the case of regulators failure. With this assumption, the system behaves as a second order nonlinear circuit, where branch inductances and the busbar capacitor dominate the eigenvalues analysis, showing a wide area of asymptotic stability. On the other hand, the droop control action was ignored, and the study appeared to be incomplete.

We moved forward with the inclusion of the droop characteristic, assuming regulators fast enough to consider that output voltage and reference voltage, calculated by droop controllers as in \eqref{eq:droop}, coincide. With this assumption---which is commonly adopted to study the interconnection of resources within a dc grid~\cite{Anand2013}---the set of conditions ensuring the asymptotic stability appears to be unfeasible.

The next step was the inclusion of the actual regulator dynamics, as in \eqref{eq:delta_alpha}. As a consequence, the system complexity increases. Nevertheless, stability can still be reached only for a set of conditions which are practically unfeasible.

Finally, we introduced a delay in the action provided by converters, by the introduction of a further state variable $\alpha_{{\rm B}_j}^{\rm ref}$, which defines the command sent to converters, to obtain the desired modulation index $\alpha_{{\rm B}_j}$.
This is useful to replicate the behavior of real components and has an essential stabilization effect, which will appear clearer in the following.

\subsection{Droop Controlled Source Equation}
The time derivative of the current of each droop-controlled source can be expressed through the \ac{kvl} of the corresponding branch circuit:
\begin{IEEEeqnarray}{rCl}\label{eq_KVL}
    \dfrac{di_{{\rm B}_j}}{dt}=\frac{1}{L_{{\rm B}_j}}\left(e_{{\rm B}_j}-R_{{\rm B}_j}i_{{\rm B}_j}-\frac{v}{\alpha_{{\rm B}_j}}\right)=f_{{\rm {i}}_j}
\end{IEEEeqnarray}
The input voltage of the dc-dc converter is derived from the main dc-bus voltage $v$, considering the modulation index of the $j$-th converter, $\alpha_{{\rm B}_j}$.

\subsection{Droop Control Equations}
The voltage regulation provided by the dc-dc converter of a droop controlled source follows a droop characteristic, driven by reference signals and source actual current, as in \eqref{eq:droop}. The regulating action is realized by the proper variation of the dc-dc converter modulation index, which is carried out using a proportional-integral regulators, fed by the reference voltage provided by the droop controller, as in \eqref{eq:delta_alpha}.

As mentioned before, the actual modulation index $\alpha_{{\rm B}_j}$ of the $j$-th converter is assumed to be obtained with a delay with respect to the reference signal $\alpha_{{\rm B}_j}^{\rm ref}$. This is represented as:
\begin{IEEEeqnarray}{rCl}\label{eq:alpha}
    \frac{d\alpha_{{\rm B}_j}}{dt}=-\frac{1}{\tau}\alpha_{{\rm B}_j}+\frac{1}{\tau}\alpha^{\rm ref}_{{\rm B}_j}=f_{{\rm \alpha}_j}
\end{IEEEeqnarray}
where $\tau$ is a time constant to be suitably tuned.

After having substituted \eqref{eq:droop} in \eqref{eq:delta_alpha}, it is possible to describe the time derivative of $\alpha^{\rm ref}$ as:
\begin{IEEEeqnarray}{rCl}\label{eq:alphacontrol}
    \dfrac{d\alpha^{\rm ref}_{{\rm B}_j}}{dt}= k_{\rm P}\left[-\dfrac{dv}{dt}-D\frac{d}{dt}\left(\frac{i_{{\rm B}_j}}{\alpha_{{\rm B}_j}}\right)\right]+\qquad\qquad\qquad\nonumber
    \\
    +k_{\rm I}\left[
   {v_0-v-D\left({\frac{i_{{\rm B}_j}}{\alpha_{{\rm B}_j}}-i_0}\right)}\right]=f_{\alpha^{\rm ref}_j}\quad
\end{IEEEeqnarray}
where, using \eqref{eq_KVL} and \eqref{eq:alpha},
\begin{IEEEeqnarray}{rCl}\label{eq_ipunto}
    \frac{d}{dt}\left(\frac{i_{{\rm B}_j}}{\alpha_{{\rm B}_j}}\right)&=&\frac{1}{\alpha_{{\rm B}_j}}\frac{di_{{\rm B}_j}}{dt}-\frac{i_{{\rm B}_j}}{\alpha^2_{{\rm B}_j}}\frac{d\alpha_{{\rm B}_j}}{dt} \nonumber \\
    &=& \frac{1}{\alpha_{{\rm B}_j}L_{\rm B_j}}\left(e_{\rm B_j}-R_{\rm B_j}i_{\rm B_j}-\frac{v}{\alpha_{{\rm B}_j}} \right)+ \nonumber\\
    & & +\frac{i_{{\rm B}_j}}{\tau \alpha_{{\rm B}_j}} - \frac{i_{{\rm B}_j}\alpha^{\rm ref}_{{\rm B}_j}}{\tau \alpha^2_{{\rm B}_j}}.
\end{IEEEeqnarray}

\subsection{Constant Power Resources}
The ideal model of a \ac{cpl} connected at the main busbar is defined as the load power  ($P_{\rm L}$), divided by the actual supply voltage. Therefore, ideal model of the CPL is nonlinear, and the load current is given by
\begin{IEEEeqnarray}{rCl}\label{eq:CPL}
    i_{\rm L}=\frac{P_{\rm L}}{v}.
\end{IEEEeqnarray}

Similarly to the \ac{cpl}, an \ac{fc}, equipped with a fast regulator designed to maintain the power production as a constant quantity, can be considered as an ideal \ac{cps}. Thus, by adopting the generators convention, the current of each \ac{fc} is \begin{IEEEeqnarray}{rCl}\label{eq:CPS}
    i_{{\rm FC}_k}=\frac{P_{{\rm FC}_k}}{v}.
\end{IEEEeqnarray}

Notice that the small-signal linearization of a \ac{cpl} or \ac{cps} around its operating point results in a negative conductance \cite{Anand2013,Su2018}.

\subsection{Busbar Current Equation}
The derivative of the busbar voltage, which corresponds to the capacitor voltage, can be expressed through the \ac{kcl} at the unique electrical node:
\begin{IEEEeqnarray}{rCl}\label{eq:KCL}
   \frac{dv}{dt}&=&\frac{1}{C}\left(\sum_j \frac{i_{{\rm B}_j}}{\alpha_{{\rm B}_j}}+\sum_k i_{{\rm FC}_k}-i_{\rm L}\right) \nonumber \\
   &=&\frac{1}{C}\left(\sum_j \frac{i_{{\rm B}_j}}{\alpha_{{\rm B}_j}}+\sum_k \frac{P_{{\rm FC}_k}}{v} - \frac{P_{\rm L}}{v}\right)=f_{\rm v}
\end{IEEEeqnarray}
\subsection{System Equations}
The equations describing the microgrid behavior can be now written as in \eqref{eq:nlinsys} with
\begin{IEEEeqnarray}{rCl}\label{eq:f}
    \boldsymbol{f} =\left[\boldsymbol{f}^T_{\rm {i}}\quad\boldsymbol{f}^T_{\rm {\alpha}}\quad\boldsymbol{f}^T_{\rm {\alpha^{ref}}}\quad f_{\rm v}\right]^T
\end{IEEEeqnarray}
\begin{IEEEeqnarray}{rCl}\label{eq:x}
    \boldsymbol{x} =\left[\boldsymbol{x}^T_{\rm {i}}\quad\boldsymbol{x}^T_{\rm {\alpha}}\quad\boldsymbol{x}^T_{\rm {\alpha^{ref}}}\quad v\right]^T
\end{IEEEeqnarray}
where
\begin{align}
    \boldsymbol{x}_{\rm {i}} &= \left[i_{{\rm B}_1}\quad i_{{\rm B}_2}\quad\cdots\quad i_{{\rm B}_n}\right]^T\label{eq:xi}\\
    \boldsymbol{x}_{\rm {\alpha}} &= \left[\alpha_{{\rm B}_1}\quad \alpha_{{\rm B}_2}\quad\cdots\quad \alpha_{{\rm B}_n}\right]^T\label{eq:xalpha}\\
    \boldsymbol{x}_{\rm {\alpha^{\rm ref}}} &= \left[\alpha^{\rm ref}_{{\rm B}_1}\quad  \alpha^{\rm ref}_{{\rm B}_2}\quad\cdots\quad \alpha^{\rm ref}_{{\rm B}_n}\right]^T.
    \label{eq:xalpharef}
\end{align}

The matrix ${\rm{\mathbf{A}}}$ of the linearized system \eqref{eq:linsys} is:
\begin{IEEEeqnarray}{rCl}\label{eq:A}
\renewcommand{\arraystretch}{1.7}
{\rm{\mathbf{A}}}=
\left.\begin{bmatrix}
J_{\boldsymbol{x}_{\rm i}} {\boldsymbol{f}}_{\rm i}&
J_{\boldsymbol{x}_{\rm \alpha}} \boldsymbol{f}_{\rm i}&
J_{\boldsymbol{x}_{\rm \alpha^{\rm ref}}} \boldsymbol{f}_{\rm i}&
J_{v}\boldsymbol{f}_{\rm i}
\\
\boldsymbol{0}_{n\times n}&
J_{\boldsymbol{x}_{\rm \alpha}} \boldsymbol{f}_{\rm \alpha}  &
J_{\boldsymbol{x}_{\rm \alpha^{\rm ref}}} \boldsymbol{f}_{\rm \alpha} &
\boldsymbol{0}_{n\times 1}
\\
J_{\boldsymbol{x}_{\rm i}} \boldsymbol{f}_{\rm \alpha^{\rm ref}} &
J_{\boldsymbol{x}_{\rm \alpha}} \boldsymbol{f}_{\rm \alpha^{\rm ref}} &
\boldsymbol{0}_{n\times n}&
J_{v} \boldsymbol{f}_{\rm \alpha^{\rm ref}}
\\
\nabla_{\boldsymbol{x}_{\rm i}} f_{\rm v} &
\nabla_{\boldsymbol{x}_{\rm \alpha}} f_{\rm v} &
\boldsymbol{0}_{1\times n}&
\frac{\partial f_{\rm v}}{\partial v}
\end{bmatrix}
\right |_{\boldsymbol{x}=\bar{\boldsymbol{x}}}
\end{IEEEeqnarray}
where $\nabla_{\boldsymbol{z}}\left(\cdot\right)$ indicate the row vector gradient operator with respect to the vector $\boldsymbol{z}$.
It is worth noting that the block elements in position (1,1), (1,2), (1,3), (2,2), (2,3), (3,1), and (3,2) in \eqref{eq:A} are diagonal matrices.

To conclude, the proposed small-signal stability analysis consists in computing matrix ${\rm{\mathbf{A}}}$, given the model parameters and an operating point $\bar{\boldsymbol{x}}$, and checking if its eigenvalues, denoted as $\lambda_i({\rm{\mathbf{A}}})$, are in the left side of the complex plane.

\section{Stability Analysis}
The small-signal stability analysis, described in Section~\ref{sec:SmallSignalModeling}, has been applied to the dc shipboard \ac{ips} depicted in Fig.~\ref{fig:Rete} with two identical \acp{ess} ($j=1,2$), and three \acp{fc} ($k=1,2,3$), whose parameters are reported in Table~\ref{tab:example_parms}. The two operating points in Table~\ref{tab:operating_points} are considered. Table~\ref{tab:example_parms} does not report the values of $L_{{\rm B}_j}$, $j=1,2$ (supposed to be equal), of $C$, and of $D$, since we want to study the sensitivity of small-signal stability on these three parameters.

\begin{table}[t]
		\centering
		\caption{Shipboard power system model parameters.}
		\label{tab:example_parms}
		{\small
		\begin{tabular}{|l|c|}
		\toprule
		\multicolumn{1}{|c|}{Parameter} & \multicolumn{1}{c|}{Value}\\
		\midrule
	    Base power $S^{\rm b}$                              &  \SI{1}{\mega\watt} \\
        Nominal voltage $V^{\rm nom}$                 &  \SI{750}{\volt} \\
        Converters control proportional gain $k_{\rm P}$    &  2 p.u. \\
        Converters control integral gain $k_{\rm I}$        &  1 p.u. \\
        \acp{ess} internal resistances $R_{{\rm B}_{j}}$   &  0.0177 p.u.\\
		\bottomrule
		\end{tabular}
        }
\end{table}

\begin{table}[t]
		\centering
		\caption{Operating points.}
		\label{tab:operating_points}
		{\small
		\begin{tabular}{|l|c|c|}
		\toprule
		\multicolumn{1}{|c|}{Parameter} & \multicolumn{1}{c|}{Operating point 1} & \multicolumn{1}{c|}{Operating point 2}\\
		\midrule
	    $v_0$                                                    &  1.000 p.u.     & 1.000 p.u.\\
	    $P_{\rm L}$                                                    &  2.950 p.u.     & 1.200 p.u. \\
        $P_{{\rm FC}_k}$                                               &  0.650 p.u.     & 0.350 p.u. \\
        $e_{{\rm B}_j}$                                                &  0.924 p.u.     & 0.935 p.u. \\
        $\bar{i}_{{\rm B}_j}$                                          &  0.546 p.u.     & 0.270 p.u.\\
        $\bar{\alpha}_{{\rm B}_j}$, $\bar{\alpha}_{{\rm B}_j}^{\rm ref}$     &  1.093 p.u.     & 1.082 p.u. \\
        $\bar{v}$                                                &  1.000 p.u.     & 1.000 p.u.  \\
		\bottomrule
		\end{tabular}
        }
\end{table}

\setcounter{figure}{3}
\begin{figure*}[t]
	\centering
    \includegraphics[width=.95\textwidth]{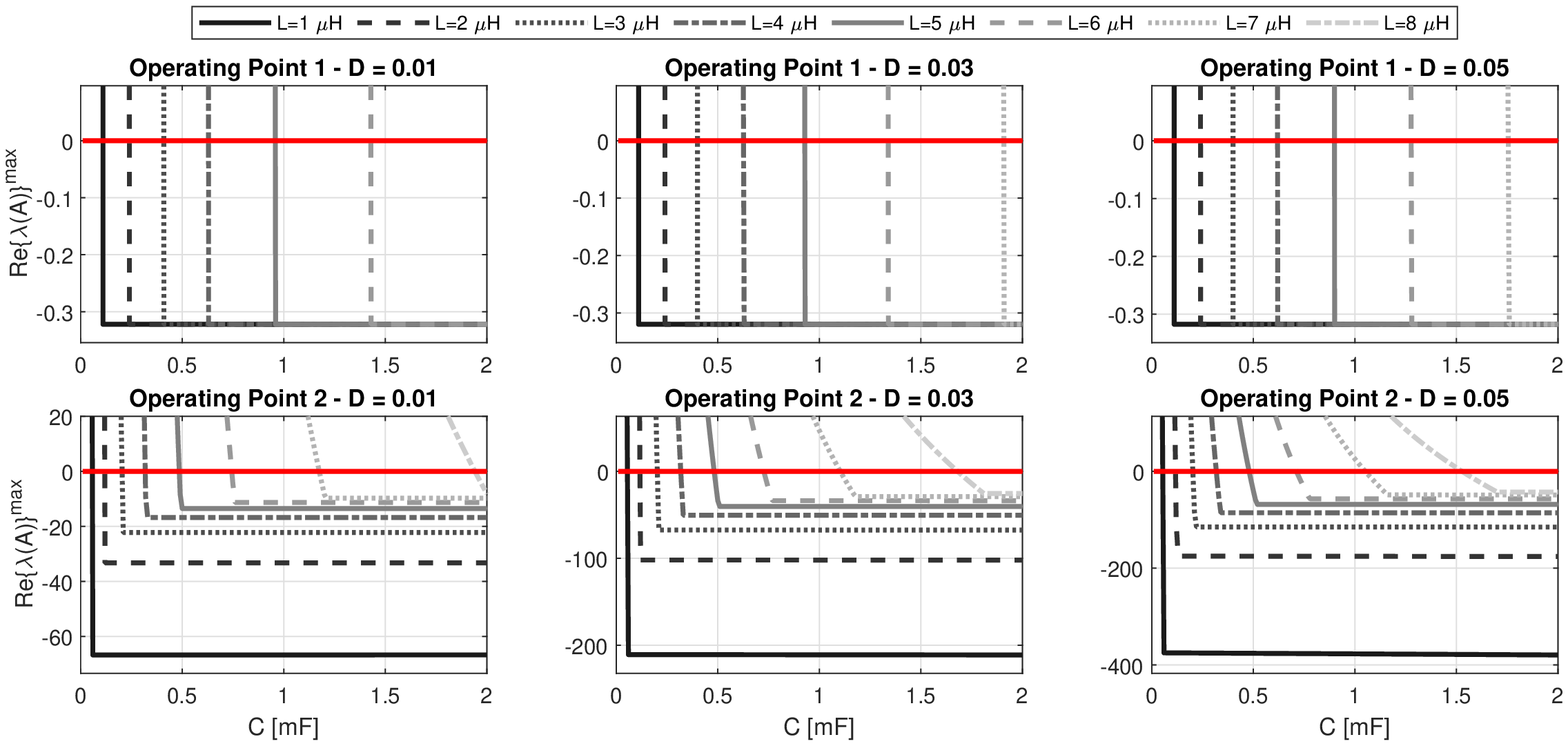}
    	\caption{Sensitivity of small-signal suitability with respect to parameters $C$, $L_{{\rm B}_j}$, and $D$.}
	\label{fig:risults1}
\end{figure*}

To prove the effectiveness of the proposed analysis, a detailed model of the \ac{ips} has been implemented in DIgSILENT Power Factory. Specifically, a DIgSILENT simulation is considered unstable if the system variables diverge merely due to numerical noise.

\setcounter{figure}{2}
\begin{figure}[t]
	\centering
    \includegraphics[width=.9\columnwidth,trim=1cm 0.1cm 1cm 0.1cm,clip=true]{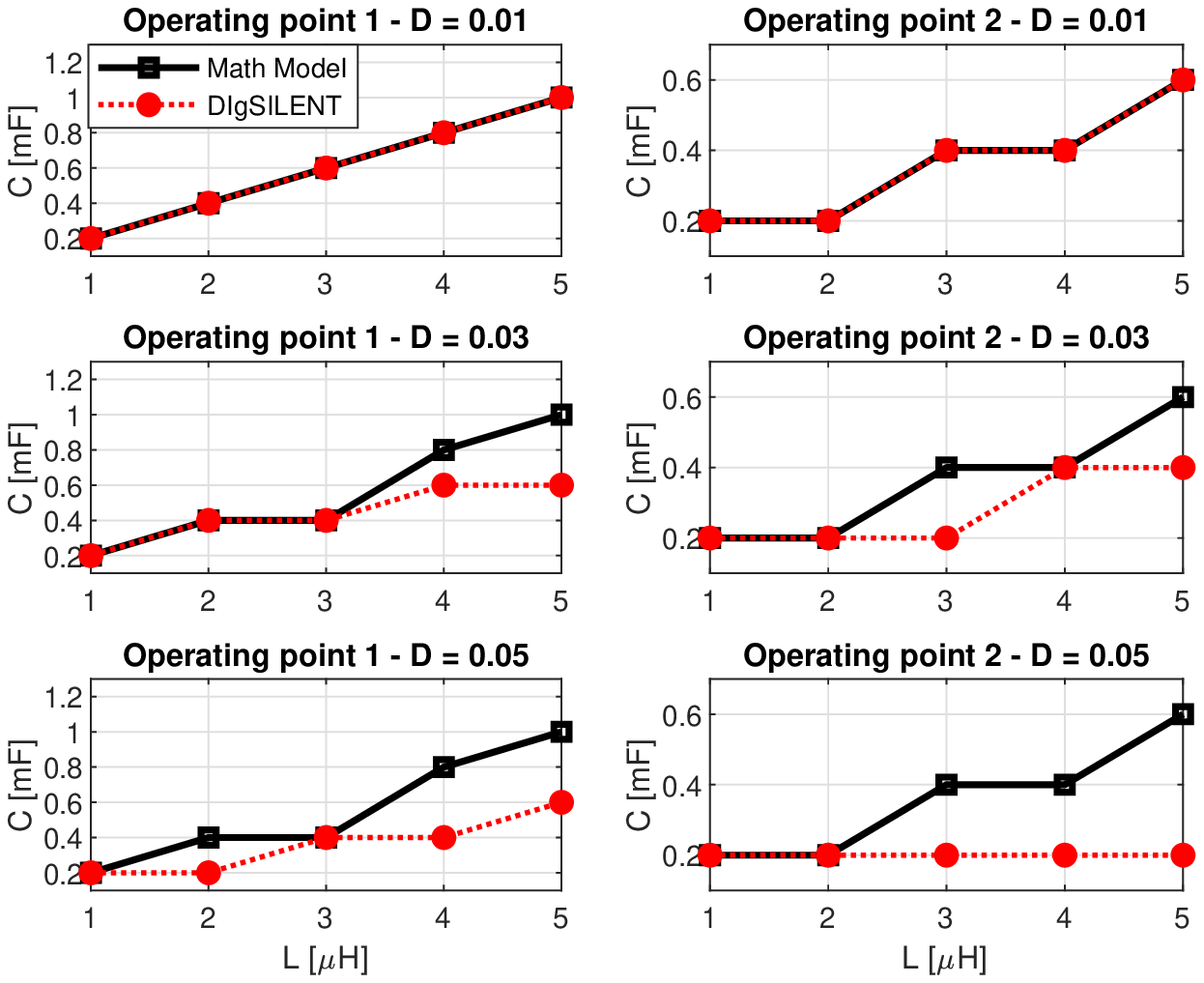}
    	\caption{Minimum capacitance $C$ to have asymptotic stability given the value of inductances $L_{{\rm B}_j}$, according to the S-SASC and DIgSILENT simulations.}
	\label{fig:risults0}
\end{figure}

A first interesting result is that by setting the converter time constant $\tau$ to very small values ($<\SI{1}{\micro\second})$, the \ac{s-sasc} $\Re\{\lambda_i({\rm{\mathbf{A}}})\}<0$) cannot be obtained with any reasonable triad of values for $C$, $D$ and $L_{{\rm B}_j}$. This is in contrast with the DIgSILENT simulations, which, within the same conditions, show asymptotic stability. Differently, larger values of $\tau$ allow obtaining the \ac{s-sasc}, with reasonable values of the three parameters. This means that the delay in the converter internal dynamics has a stabilizing effect. Obviously, $\tau$ has to be small enough to suitably represent the converter internal dynamics, i.e., $\tau\leq \SI{1}{\milli\second}$.

We performed a tuning procedure looking for the value of $\tau$ to obtain the matching between the theoretical \ac{s-sasc} and the asymptotic stability of the DIgSILENT simulations. Since the converter internal dynamics is only approximated by \eqref{eq:alpha}, a perfect matching was impossible to be found. Thus, a tuning procedure has been carried out looking for the minimum $\tau$ such that DIgSILENT simulations result to be asymptotic stable when the theoretical \ac{s-sasc} is satisfied. Such optimal value has resulted to be $\tau^*=\SI{0.9}{\milli\second}$. In this way, the \ac{s-sasc} has become sufficient but not necessary.

One of the interesting aspects in analyzing the system stability concerns the sizing of $C$ and $L_{{\rm B}_j}$. Under the hypothesis that the dc-bus capacitor has a stabilizing effect on the system, in Figure~\ref{fig:risults0} we show the minimum capacitance $C$ satisfying the \ac{s-sasc}, with a resolution of \SI{0.2}{\milli\farad}, given the value of inductances $L_{{\rm B}_j}$, for both the operating conditions and for three values of the droop gain $D$. In Figure~\ref{fig:risults0}, such values are also compared with the ones obtained with the DIgSILENT simulations. We can observe that the DIgSILENT values are lower than the theoretical one, confirming the sufficiency, and therefore the robustness, of the \ac{s-sasc}.

A further fact that we can observe in Figure~\ref{fig:risults0} is that larger values of inductances $L_{{\rm B}_j}$ and smaller values of the droop coefficient $D$ make the system less stable, since higher values of $C$ are required to guarantee the asymptotic stability.

The sensitivity of the small-signal asymptotic stability on the values of parameters $C$, $L_{{\rm B}_j}$, and $D$ is further investigated using the \ac{s-sasc} as showed in Figure~\ref{fig:risults1}.\balance\@ In particular, this figure shows the maximum real part of the eigenvalues of matrix ${\rm{\mathbf{A}}}$, $r^{\rm max} = \max_i \Re \{\lambda_i({\rm{\mathbf{A}}})\}$, in both the operating conditions and with three values for $D$. We can observe that, given a fixed value for $L_{{\rm B}_j}$, increasing $C$ from values lower than $\SI{0.1}{\milli\farad}$, $r^{\rm max}$ moves from positive to negative values, confirming the stability effect of a larger capacitance. Moreover, we observe that, the larger the inductances $L_{{\rm B}_j}$, the larger the minimum $C$ required to get asymptotic stability. This confirms the conjecture deduced by analysing results in Figure~\ref{fig:risults0}. The same can be said about the sensitivity on $D$.

\section{Conclusion}\label{sec:Conclusion}
This work presented a stability analysis of a dc shipboard \ac{ips} made up of \acp{fc}---modeled as \acp{cps}---, \acp{cpl}, and \acp{ess} equipped with droop controllers. In particular, the duty cycles of the \acp{ess} have been included in the mathematical model describing the system, and a design criterion has been derived. This criterion provides the values for droop, dc-bus capacitance, and dc-dc converters inductances sufficient to attain the stability of the system.



\end{document}